\documentclass[a4paper,twocolumn,11pt]{article}
\usepackage[spanish]{babel}
\usepackage{ae}
\usepackage[latin1]{inputenc} 
\usepackage[T1]{fontenc} 
\usepackage{layout}
\usepackage{array}
\usepackage{graphicx}
\usepackage{amsmath}
\usepackage{multicol}
\hyphenrules{nohyphenation} 

\usepackage[]{cite}   
\usepackage{anysize}  
\usepackage{endnotes} 

\usepackage{tikz,fp,ifthen,fullpage}
\usetikzlibrary{arrows,snakes,backgrounds}
\usepackage{pgfplots}
\usetikzlibrary{calc,through,backgrounds,decorations}
\usetikzlibrary{decorations.shapes,decorations.text,decorations.pathmorphing,backgrounds,fit,calc,through,decorations.fractals}
\usetikzlibrary{fadings,intersections}
\usetikzlibrary{patterns}
\usetikzlibrary{mindmap}

\marginsize{1.5cm}{1.0cm}{1.5cm}{2cm}
\begin{document}

\renewcommand{\tablename}{Tabla}
\renewcommand{\abstractname}{}
\renewcommand{\thefootnote}{\arabic{footnote}}

\title{
        {\LARGE Diagrama semi-cuantitativo sobre evolución temporal del aprendizaje de la física elemental: un estudio con alumnos de ingeniería}\\       
\footnotesize   \textit{(Semi-quantitative diagram about temporal evolution in basic physics learning: a study with engineering students)}
}
 
\author
{ 
Paco Talero $^{1,2}$, César Mora $^{2}$\\
Orlando Organista $^{1,2}$, Luis Barbosa $^{1,2}$,\\
$^{1}$ {\small Grupo F\'{\i}sica y Matem\'{a}ticas, Depto de Ciencias
Naturales, Universidad Central},\\ 
\small {Carrera 5 No 21-38, Bogot\'{a}, D.C.Colombia.}, \\ 
$^{2}$ {\small Facultad Centro de Investigaci\'{o}n en Ciencia Aplicada y
Tecnolog\'{\i}a Avanzada del Instituto Polit\'{e}cnico Nacional,} \\
\small {Av. Legaria
694, Col. Irrigaci\'{o}n, C. P. 11500, M\'{e}xico D. F.}
}
\date{}
\twocolumn
[
\begin{@twocolumnfalse}
\maketitle 
\begin{abstract}
Se presenta una técnica novedosa de análisis semi-cuantitativo plasmada en un diagrama geométrico que permite estudiar algunas características de la evolución temporal del aprendizaje de conceptos concretos de física elemental durante periodos sucesivos de tiempo. Esta técnica cuantifica el grado de homogeneidad $(H)$ y el rendimiento $(S)$ de la población en  periodos sucesivos de tiempo que son organizados en un diagrama $(H-S)$ que contiene información cuantitativa y cualitativa sobre las características  del aprendizaje y la instrucción impartida. Esta técnica fue aplicada al estudio de la evolución temporal de la interpretación gráfica del  movimiento uniforme rectilíneo, a través de tutoriales en un curso de física introductoria con $20$ alumnos de ingeniería en  la Universidad Central de Bogotá Colombia durante un periodo de  $6$ semanas. Se encontró en el diagrama global que $(Smax,Hmax)\approx (0.75,0.75)$.\\
\textbf{Palavras-chave:} diagrama, evolución temporal, enseñanza de la física, métodos de enseñanza, homogeneidad. \\ 

We show a new semi-quantitative technique of analysis about of temporal evolution of learning of basic physics concepts. In this technique we arrange a geometric diagram with the score  and the homogeneity $(H-S)$, in this diagram we have a quantitative and qualitative information about the learning   of students and the effectiveness instruction. We applied this technique to study the temporal evolution of the graphical interpretation of motion in one dimension    through tutorials in a introductory course physics with $20$  students of engineering at the Universidad Central de Bogotá Colombia for a period of $6$  weeks. We found in the general diagram that $(Smax,Hmax)\approx (0.75,0.75)$.\\
\textbf{Keywords:} diagram, evolution, physics education, teaching methods, homogeneity.\\
\end{abstract}
\end{@twocolumnfalse}
]

\section{Introducción}
Durante los últimos treinta años múltiples investigaciones  en el campo de la enseñanza y aprendizaje de la física, han venido mostrando una serie de \-pro\-ble\-mas fundamentales comunes a diferentes poblaciones de un mismo país e incluso comunes a diferentes países \cite{LieB_1,LieB_2,LMC1,LieB_3,Esp}, entre los problemas más destacados se encuentran: primero que las ideas naturales, pre-concepciones, ideas erróneas o ideas alternativas que tienen los estudiantes no son removidos en la educación básica; segundo que los estudiantes continúan con las mismas ideas previas aún después de tomar sus asignaturas de física general en la educación superior; tercero que los estudiantes usan modelos mentales similares y erróneos a la hora de explicar un fenómeno físico y cuarto que la enseñanza, en general, no tiene en cuenta las concepciones de los estudiantes ni sus características siendo de esta manera poco efectiva a la hora de formar ciudadanos con sólidos conocimientos básicos de física\cite{LieB_1,LieB_2}. 

El reconocimiento de estos problemas ha estimulado la formación de comunidades científicas, como la Physics Education Research (PER) que busca enfrentar y proponer soluciones a este tipo de  problemas \cite{LMC1}. Así se han concebido diversas alternativas de instrucción física como los tutoriales, instrucción por pares y experimentos discrepantes \cite{LMC2,Julio,Botellas,Globo}, entre muchas otras técnicas de intrucción enmarcadas en el aprendizaje activo de la física\cite{F_ACT}. Por otra parte, las dinámicas de investigación cuantitativa en la enseñanza de la física han permitido construir  diversos  instrumentos de investigación como el FCI, TUG-K y BEMA   entre otros \cite{Hestenes1,Beichner,BEMA}, que  son test  de selección múltiple cuyas preguntas tienen distractores que sintetizan las tendencias del razonamiento común y evidencian la presencia de modelos mentales intuitivos. 
 
La utilidad de estos instrumentos de indagación ha sido potenciada con la construcción  de técnicas cuantitativas que permiten organizar los datos obtenidos y ayudan a obtener información. En la actualidad dos técnicas predominan en la literatura de investigación cuantitativa  de la enseñanza de la física: el factor de concentración y la ganancia media normalizada \cite{Bao2,Hake}. El factor de concentración permite medir la concentración $C$ de las respuestas en las alternativas de una pregunta de selección múltiple y relacionarlas con el rendimiento o puntaje $S$, esta información es usada para conocer, entre otras cosas, si los estudiantes tienen modelos incorrectos o no. La ganancia media normalizada $G$ es comúnmente usada en metodologías que evalúan mediante pre y pos-test,  mide la diferencia de puntaje  entre estas pruebas, tomando como base el puntaje del pre-test $S_o$ y del pos-test $S_f$. Así mismo, en\cite{Quepasa} se encuentra un estudio realizado sobre física introductoria que permitió  caracterizar la evolución de la comprensión  de los estudiantes de manera más detallada de lo que es posible a través de un pre y  post-test. Las curvas de rendimiento $S-t$ encontradas  muestran que en algunos aspectos no hay cambios significativos, en otros hay decadencia, oscilación e incluso hay tendencia a la baja en algunos ítems.

De otro lado, un enfoque multidisciplinar basado en los conceptos de la sociología, psicología educativa, física estadística y ciencias de la computación ha venido desarrollando descripciones teóricas de los procesos de enseñanza-aprendizaje en el aula de clase con buenos resultados\cite{Alvano,Redish1,Lee,Nitta}. Así se ha estudiado mediante modelos teóricos los resultados brindados por metodologías de instrucción como el aprendizaje colaborativo, aprendizaje social a través de internet e  instrucción  por pares\cite{Nitta,10Anos}, entre otras. Estas ideas teóricas han explorado tanto el aprendizaje como las metodologías de instrucción y se han realizado propuestas analíticas y desarrollos mediante simulación.
 
Pese al gran desarrollo de la investigación  en enseñanza de la física, no se ha reportado en la literatura la construcción de una técnica  de análisis que dé cuenta de algunas características de la evolución temporal del aprendizaje de conceptos de física elemental durante periodos sucesivos de tiempo. En este trabajo se presenta una propuesta que implementa tal técnica de análisis, para esto en la sección $2$ se presentan las definiciones fundamentales que delimitan el trabajo; en la sección $3$ se exponen los criterios que permiten cuantificar el concepto de homogeneidad a partir de test de selección múltiple; en la sección $4$ se definen y ejemplifican los diagramas de evolución del aprendizaje de la física (DEAF); en la sección $5$ se presentan los resultados de una investigación realizada en la facultad de ingeniería de la Universidad Central de Bogotá con estudiantes de ingeniería donde se aplicó DEAF y finalmente en la sección $6$ se presentan las conclusiones. 

\section{Definición fundamentales}

Las características generales de los procesos de enseñanza aprendizaje de la física son en general  muy  complejas y abarcan  dinámicas propias a las instituciones, las directivas, así como de  los  profesores  y  los estudiantes con todas sus características psicológicas y socio-culturales. En cuanto a los estudiantes se refiere se sabe que debido a su pasado académico, experiencia de vida individual, predisposición, motivación, creatividad, tiempo de trabajo y organización poseen un ritmo de trabajo  que determina el desempeño a lo largo del proceso de enseñanza aprendizaje. Lo anterior  implica que en general resulta muy complejo  de evaluar y cuantificar este tipo de procesos. Sin embargo, múltiples estudios realizados por la PER  muestran que hay algunas regularidades en este tipo de procesos que requieren atención e investigación \cite{LMC1}.  

De acuerdo con lo anterior, se define  el sistema en estudio como un conjunto de $N_e$ ($N_e>2$) estudiantes que están dispuestos en una u otra  medida a aprender física, se reúnen periódicamente durante algún  tiempo para interactuar entre ellos y con el  profesor quien es el responsable de asignar, orientar y dirigir  las diversas actividades académicas en el marco de  una metodología propia, pero dentro de un referente curricular claro, preciso y en armonía con los objetivos de una institución. Así, se entiende el sistema de estudio compuesto por estudiantes,  profesor, metodología e institución y  el objeto de estudio es el aprendizaje  de la física del grupo de estudiantes y la efectividad de las instrucciones.

Llámese observable de aprendizaje a toda propiedad o característica asociada con el aprendizaje de los estudiantes susceptible de ser medida con plena claridad y precisión. Cada sistema se caracteriza por un conjunto bien definido de observables de aprendizaje, el desempeño o puntaje $S$, la concentración $C$ y la ganancia media normalizada $G$ son algunos  ejemplos.

Además, para materializar los observables de aprendizaje se requiere construir un test de indagación que cumpla preferiblemente con las siguientes características: 

\begin{enumerate}
	\item  surgir de las peculiaridades de la población en estudio con sus ideas  
	      erróneas y tendencias de pensamiento común, que se pueden obtener a partir de 	  
		 entrevistas directas o conocimientos extraídos de otras investigaciones.
	\item  ponerse a punto con una muestra de la misma población que no participa en el estudio, allí 
	 se analiza  la conveniencia de cada pregunta en cuanto a redacción y pertinencia 
	 disciplinar se refiere.
	\item tener $m$ opciones y por tanto $m-1$ distractores que  inscriban las ideas 
	   alternativas y erróneas características de esta población. 
	\item contener $N_p$ preguntas que apuntan a indagar diferentes matices de un concepto o pensamiento 
	     físico particular, por ejemplo el MUR.
\end{enumerate}

Si un test como el descrito anteriormente se aplica a $N_e$ estudiantes el resultado de la prueba se puede representar mediante la tabla \ref{tab1}, donde el número de opciones  $1,2,3,\cdots, m$  se ha denotado con  la siguiente asignación de correspondencias $1\rightarrow A , 2\rightarrow B,\ldots$ etc.; los estudiantes participantes  se denotan como $e_1,e_2,e_3\ldots e_i\ldots e_{N_e}$ y las preguntas del test se denotan como $p_1,p_2\ldots p_j \ldots p_{N_p}$, esta configuración  se llamará en adelante el estado del sistema. 	

\begin{table}[htp]

\caption{Estado del sistema}\label{tab1} 
\vspace{2mm}

\centering
\begin{tabular}{|c||c|c|c|c|c|c|c|}
 \hline  
  
 \textsf{} & $p_1$  & $p_2$  & $p_3$  & $\ldots$  & $p_j$  & $\ldots$  & $p_{N_p}$  \\ \hline  \hline

 $e_1$      & A  & E & D &C & E & E& B \\ \hline  
 $e_2$      & C  & $\ldots$ &  & & & & $\vdots$  \\ \hline  
 $e_3$      & D  & $\ldots$ &  & & & & $\vdots$    \\ \hline 
 $\vdots $  & C  & $\ldots$  &  & $\ddots $ & & & $\vdots$   \\ \hline  
 $e_i$      & E  &  $\ldots$ &  & & & & $\vdots$   \\ \hline  
 $\vdots$   & B  &  $\ldots$ &  & & & & $\vdots$    \\ \hline  
 $e_{N_e}$  & A  & $\ldots$ &  & & & &  \\ \hline    
 
\end{tabular}
\end{table}

\section{Homogeneidad}

Para caracterizar la evolución temporal del pensamiento de un grupo en un ámbito particular de la física es preciso definir la homogeneidad respecto a una pregunta o grupo de preguntas  que responda o respondan a los diferentes matices de pensamiento físico a estudiar. El análisis de concentración \cite{Bao2} permite cuantificar la distribución de la concentración de una pregunta en particular, pero no se conocen, en cuanto a este campo de investigación se refiere, técnicas que permitan  cuantificar  la homogeneidad en una pregunta o de manera global. En esta sección se muestra  como construir un indicador que cuantifique el grado de homogeneidad de una pregunta y de la prueba a nivel global. 

Un grupo de  individuos se considera  homogéneo cuando sus miembros comparten  características en un contexto determinado. En la investigación educativa  generalmente el grupo homogéneo en estudio corresponde a un grupo de estudiantes con características similares en contextos tan diversos como el cultural, la condición socio-económica, el sexo, la experiencia de vida, los valores y el  rendimiento académico, entre muchos otros. 

Por otra parte, uno de los resultados más importantes de la investigación en enseñanza de la física es que los estudiantes al iniciar un curso ya traen consigo diferentes ideas erróneas, por lo general  difíciles de cambiar, que suelen  activarse a la hora de explicar diversos aspectos de un mismo hecho físico, también se sabe que  en general la gama de estas concepciones alternativas en una población  particular tiende a ser pequeña \cite{Bao1}. Esto ha propiciado que tales ideas alternativas  sean  utilizadas por muchos investigadores para  construir  instrumentos de opción múltiple que utilicen  las  concepciones alternas  reportadas por los estudiantes  para construir distractores  que tengan  sentido para el instructor\cite{Bao1}. Se desprende de lo anterior que un grupo de estudiantes de física con un entorno soci-cultural bien delimitado  resulta homogéneo en alguna medida ya que las investigaciones  reportan en su mayoría tendencias de pensamiento común\cite{LMC1}. Si se toma como criterio de homogeneidad la coincidencia en las ideas físicas erróneas o no que tienen los estudiantes entonces surge  la necesidad de cuantificar el  concepto de homogeneidad de una estructura global de pensamiento físico, por ejemplo el pensamiento Newtoniano- o más puntualmente: el pensamiento físico de los estudiantes en torno al MUR-\cite{Hestenes1}.

Se define el número de coincidencias $N_c$, como el número total de coincidencias que los estudiantes tienen al responder cada opción en cada pregunta. Así, sea $n_i$ el número de coincidencias que tiene el estudiante $i$-ésimo con los estudiantes $e_{i+1},e_{i+2} \ldots  e_{i+N_e}$, de manera que el núnero total de coincidencias está dado por la suma del número de coincidencias  $n_1,n_2,n_3,n_i\ldots n_{N_e-1} $, es decir:

\begin{equation}\label{ecua1}
  N_{c}= \displaystyle\sum_{i=1}^{N_{e}-1} n_{i}.
\end{equation}

Ahora, para construir un indicador de homogeneidad es preciso encontrar el número máximo ($N_{cmax}$) de coincidencias y el mínimo número de coincidencias ($N_{cmin}$), para ello se procede de manera siguiente.

El máximo número de coincidencias ocurre cuando en la configuración de respuesta todos los estudiantes escogen la misma opción, de acuerdo con esto el número de coincidencias del   $i$- ésimo estudiante es   $n_i=N_p (N_{e}-i)$ y usando la expresión (\ref{ecua1}) se encuentra

\begin{equation}\label{ecua2}
  N_{cmax}=\frac{1}{2} N_{p} N_{e} (N_{e}-1).
\end{equation}

Para encontrar una configuración con el mínimo número de coincidencias es necesario considerar dos situaciones: la primera cuando el número de estudiantes es múltiplo del número de opciones de cada pregunta del test, es decir $N_{p}=mN_{e}$ y la segunda cuando no lo es.

En el primer caso se pueden disponer diferentes opciones en cada pregunta hasta agotar las posibilidades formando bloques en los cuales no hay coincidencia interna al bloque. El número de coincidencias se obtiene al notar que un estudiante en particular tiene una coincidencia por cada bloque, dado que se tienen $N_{b}-1$ bloques, $m$ opciones y $N_p$ preguntas el número de coincidencias mínimo en este caso es $m(N_{p}-1)N_p$. En el segundo caso es preciso sumar todas las coincidencias que tiene cada estudiante con el bloque incompleto,  puesto que  adicionalmente hay $z=N_{e} \mod{m} $ elementos en el bloque sobrante y el número de bloques completos son $N_{b}=\left\lfloor  \frac{N_e}{m}\right\rfloor$ entonces el número adicional de coincidencias por pregunta es $zN_b$.

De acuerdo con lo anterior el número mínimo de coincidencias está dado por:

\begin{equation}\label{ecua3}
N_{cmin}= N_{p} \left[m(N_{b}-1)+zN_{b} \right].
\end{equation}

Se define el coeficiente de homogeneidad como un número real perteneciente al intervalo $[0,1]$ que indica el parecido entre todos los estudiantes en cuanto al ítem elegido se refiere. $H=1$ significa que son completamente iguales mientras que $H=0$ significa que son completamente diferentes, es decir lo más diferente que pueden ser de acuerdo con las características  del test. La relación entre el número de coincidencias y la homogeneidad permite inferir la evolución del sistema con el transcurso del tiempo, si la evolución se presenta de manera que los elementos $(N_c,H)$ de esta relación  tomen valores que hagan mínima la distancia $(N_{cmin},0)\rightarrow (N_{cmax},1)$  se tendrá el trayecto de evolución con mejor desempeño metodológico. De acuerdo con lo anterior, si se define una función ideal idéntica entre $N_c$ y $H$  exigiendo las condiciones $H=0$ cuando $N_c=N_{cmin}$ y $H=1$ cuando  $N_{c}=N_{cmax}$  se encuentra

\begin{equation}\label{ecua4}
H=\frac{ N_{c}-N_{cmin} }{ N_{cmax}-N_{cmin} }.
\end{equation}

La expresión (\ref{ecua4}) permite cuantificar la homogeneidad del estado del sistema en estudio, en particular permite estudiar la homogeneidad del estado de una única pregunta haciendo $N_{p}=1$.  

\section{Diagrama de evolución temporal}

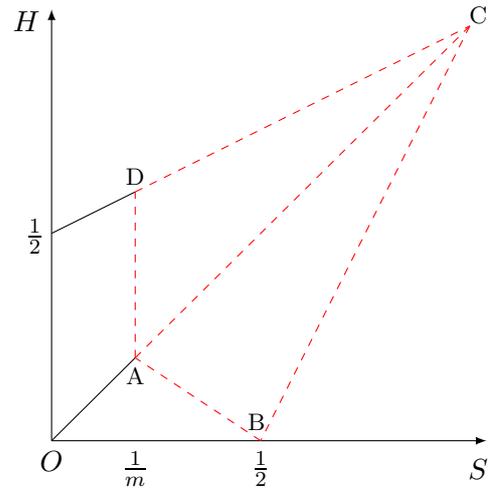
\begin{figure}[ht]
\begin{center}
\begin{tikzpicture}[scale=1.1,decoration=Koch snowflake]
       \draw [arrows=-latex] (0,0)--(5.2,0);
       \draw [arrows=-latex] (0,0)--(0,5.2);
       \coordinate [label=below:\textcolor{black} {$S$}] (x3) at (5.1,-0.1cm);
       \coordinate [label=below:\textcolor{black} {$H$}] (x3) at (-0.3,5.3cm);
       \coordinate [label=below:\textcolor{black} {$O$}] (x3) at (0.0,0.0cm);
     \coordinate [label=below:\textcolor{black} {$\frac{1}{2}$}] (x3) at (2.5,0.0cm);
     \coordinate [label=below:\textcolor{black} {$\frac{1}{2}$}] (x3) at (-0.2,2.8cm);
     \draw[thin] (0,2.5)--(1,3);
     \draw[thin] (0,0)--(1,1);
     \draw[dashed,red] (1,1)--(1,3);
     \draw[dashed,red] (1,3)--(5,5);
     \draw[dashed,red] (5,5)--(2.5,0);
     \draw[dashed,red] (2.5,0)--(1,1);
	\draw[dashed,red] (1,1)--(5,5);
     \coordinate [label=below:\textcolor{black} {\footnotesize{A}}] (x3) at (1,1cm);
     \coordinate [label=below:\textcolor{black} {\footnotesize{B}}] (x3) at (2.45,0.45cm);
     \coordinate [label=below:\textcolor{black} {\footnotesize{C}}] (x3) at (5.1,5.35cm);
     \coordinate [label=below:\textcolor{black} {\footnotesize{D}}] (x3) at (1.0,3.4cm);
     \coordinate [label=below:\textcolor{black} {$\frac{1}{m}$}] (x3) at (1.0,0.0cm);

\end{tikzpicture}
\caption{-El diagrama DEAF.}
\label{deaf}
\end{center}
\end{figure}

El estudio de la evolución del aprendizaje de conceptos concretos de física elemental que se propone en este trabajo consiste en la aplicación consecutiva de un test durante sucesivas secciones de clase, para indagar por la evolución de los aciertos y falencias disciplinares de los estudiantes y la metodología de instrucción. De esta manera se   observa  la evolución en tiempo de ejecución, desde un punto de vista cuantitativo y cualitativo. Para esto se calcula tanto el desempeño $S$ como la homogeneidad $H$ del estado instantáneo del sistema en cada sección y se organiza en un diagrama $H-S$ como el mostrado en la Fig.\ref{deaf}. Además,  esta información es complementada con la extraída de un diario de campo que lleva el profesor. De acuerdo con lo anterior el profesor tiene suficiente información para ir ajustando su metodología de instrucción y observar de manera inmediata los resultados de la intervención.   

Para construir un diagrama semi-cuantitativo sobre evolución temporal del aprendizaje de la física elemental DEAF se asigna al eje horizontal el desempeño o puntaje $S$ normalizado a $1$ y al eje vertical se le asigna el valor de homogeneidad $H$. Cuando el puntaje es cercano a $\frac{1}{m}$, siendo $m$ el número de opciones, hay un predominio del azar en la prueba y por lo tanto no se espera puntajes por debajo de   $\frac{1}{m}$. Cuando $S\approx\frac{1}{2}$ se tiene máxima incertidumbre en el desempeño, de manera que puntajes mayores a $\frac{1}{2}$ muestran una tendencia a la  apropiación del concepto físico, mientras que  puntajes  menores a $\frac{1}{2}$ muestran una prueba más bien gobernada por el azar. De manera similar $H>\frac{1}{2}$ marca una tendencia hacia lo homogéneo o parecido, mientras que  $H<\frac{1}{2}$ indica una diferencia en la percepción tanto de los conceptos correctos como de  ideas y alternativas erroneas.     

Un punto determinado $(S,H)$ del DEAF cuantifica el estado del sistema, de manera que la distribución de cada punto en este diagrama muestra la evolución temporal del sistema en estudio. En la Fig.\ref{deaf}  la trayectoria cerrada $ABCDA$ indica la zona más significativa en cuanto a evolución coherente se refiere, pues zonas con $\frac{1}{m}$   serán visitadas con poca frecuencia ya que se parte de un test debidamente construido; la zona por encima de la semi-recta $\overline{DC}$ indica una alta homogeneidad la cual no se espera frecuentemente cuando hay bajo puntaje, ni se espera que la homogeneidad sea muy alta con puntaje medio; a la derecha de la semi-recta $\overline{BC}$  no se esperan visitas, pues el aumento del desempeño implica  un aumento en la homogeneidad; la zona por debajo de la semi-recta $\overline{AB}$ es de alta aleatoriedad ya que tiene un puntaje bajo y es altamente heterogénea; la zona alrededor del punto $A$ indica un puntaje de bastante azar y de baja homogeneidad lo que significa un punto de partida muy probable para el sistema, mientras que los alrededores del punto $C$ indican un máximo tanto de desempeño como de homogeneidad y  la semi-recta $\overline{AC}$ indica el trayecto más eficiente de aprendizaje, puesto que es un camino de pocos intentos para alcanzar alto puntaje, es decir alta homogeneidad y por tanto una metodología exitosa. De acuerdo con lo anterior  es innecesario usar todo el armazón gráfico de la Fig.\ref{deaf}, basta con tomar la zona delimitada con los puntos $ABCDA$ y mantener su orientación, este diagrama se denotará  en adelante como DEAF.    

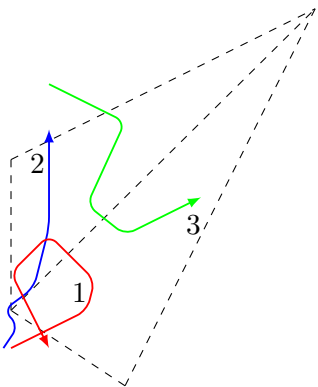
\begin{figure}[ht]
\begin{center}
\begin{tikzpicture}[scale=1.0,decoration=Koch snowflake,x=1cm,y=1cm]
     \draw[dashed,black] (1,1)--(1,3);
     \draw[dashed,black] (1,3)--(5,5);
     \draw[dashed,black] (5,5)--(2.5,0);
     \draw[dashed,black] (2.5,0)--(1,1);
	\draw[dashed,black] (1,1)--(5,5);
     \draw[-latex,rounded corners,blue,line   
           width=0.7pt](0.9,0.5)--(1.1,0.8)--(0.9,1)--(1.3,1.3)--(1.5,2.1)--(1.5,3.4);
     \draw[-latex,rounded corners,green,line 
           width=0.7pt](1.5,4.0)--(2.5,3.5)--(2.0,2.4)--(2.5,2)--(3.5,2.5); 
     \draw[-latex,rounded corners,red,line 
           width=0.7pt](1,0.5)--(2,1)--(2.1,1.4)--(1.5,2)--(1,1.5)--(1.5,0.5); 
      \coordinate [label=below:\textcolor{black} {$1$}]   (x3) at (1.9,1.5cm);
      \coordinate [label=below:\textcolor{black} {$2$}]   (x3) at (1.35,3.2cm);
      \coordinate [label=below:\textcolor{black} {$3$}]   (x3) at (3.4,2.4cm);
          
\end{tikzpicture}
\caption{-Algunos posibles trayectos de evolución.}
\label{Trayec}
\end{center}
\end{figure}	

Diversas trayectorias de evolución del sistema se pueden presentar: en la Fig.\ref{Trayec} se ilustran algunos ejemplos típicos: el trayecto $(1)$ muestra un ciclo cercano al punto $A$ y por tanto indica una metodología muy poco eficiente en la cual los estudiantes, respecto a sus conocimiento disciplinares,  salen del curso prácticamente igual a como entraron; el trayecto $(2)$ indica un escenario en el que el sistema inicia en un estado de baja homogeneidad y bajo puntaje y evoluciona hacia un estado de alta homogeneidad y bajo puntaje, tal situación eventualmente podría darse si se va creando poco a poco creencia en una o varias ideas erróneas y el trayecto $(3)$ muestra como partiendo de una alta homogeneidad dada por existencia de errores comunes muy arraigados el sistema evoluciona hacia una zona de menor homogeneidad pero mayor puntaje, dando a entender que hay cierta efectividad en la metodología de instrucción  usada.  

El DEAF permite estudiar el comportamiento global o particular de un sistema en cuanto al aprendizaje disciplinar de conceptos concretos de física elemental se refiere. Para estudiar un concepto determinado se puede usar el DEAF aplicado a una única pregunta y si se quiere estudiar todo un pensamiento físico se usa la prueba completa. 

\section{Aplicación del DEAF a estudiantes de Ingeniería en la UC}

La facultad de ingeniería de la Universidad Central está ubicada en la zona centro de Bogotá, esta facultad tiene inscritos  alrededor de $3500$ estudiantes de diversas ingenierías que toman cursos de física I, física II y física III, por lo general en grupos de $20$ estudiantes durante cuatro horas semanales. En el primer curso se abordan temas de mecánica Newtoniana de la partícula desde un punto de vista teórico-experimental con uso de una metodología mayoritariamente magistral,  tomando como prerrequisito el algebra lineal y como correquisito el cálculo diferencial de una variable.  

Durante el segundo semestre del año 2011 se realizó un estudio  con $180$ estudiantes que buscaba indagar la efectividad al interpretar gráficamente el movimiento rectilíneo. Para esto se aplico el test TUG-K a estudiantes de física I después de haber tomado el curso bajo acción metodológica magistral caracterizada por la exposición del profesor y la actuación de los estudiantes sólo a través de ejercicios, laboratorios y exámenes. Los resultaron mostraron una distribución prácticamente binomial con probabilidad del orden de $\frac{1}{5}$, lo que mostró la poca efectividad de la instrucción tradicional en esta institución.

Durante el primer semestre de $2012$ se investigó la evolución del pensamiento físico del MUR en estudiantes de física I, con la hipótesis de trabajo siguiente: si  un curso de 20 estudiantes se somete  a una instrucción activa guiada por tutoriales se tendrá una evolución que iría desde  zonas de bajo rendimiento y baja homogeneidad hasta alcanzar un alto desempeño y alta homogeneidad.

Para llevar a cabo esta investigación se diseñó y puso a punto un test de $8$ preguntas de selección múltiple con única respuesta, que indagaba por  los conceptos de posición, velocidad y desplazamiento en el MUR, enmarcado dentro de la interpretación gráfica. También, se aplicó un tutorial sobre este tema diseñado para materializar esta metodología. El test se aplicó en cada sesión de clase durante $6$ semanas, en total se realizaron $10$ aplicaciones consecutivas.

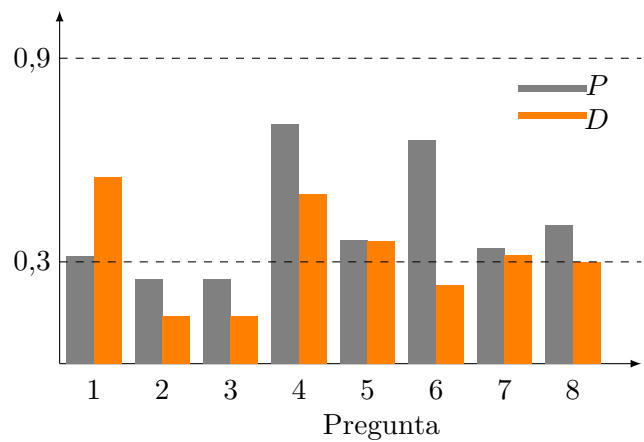
\begin{figure}[ht]
\begin{center}
\begin{tikzpicture}[scale=0.9,decoration=Koch snowflake,x=1cm,y=1cm]
       \draw [arrows=-latex] (0.5,0)--(9.0,0);
       \draw [arrows=-latex] (0.5,0)--(0.5,5.2);
       \coordinate [label=below:\textcolor{black} {Pregunta}] (x3) at (5.2,-0.6cm);      
       \coordinate [label=below:\textcolor{black} {$1$}] (x3) at (1,-0.1cm);
	  \coordinate [label=below:\textcolor{black} {$2$}] (x3) at (2,-0.1cm);
	  \coordinate [label=below:\textcolor{black} {$3$}] (x3) at (3,-0.1cm);
	  \coordinate [label=below:\textcolor{black} {$4$}] (x3) at (4,-0.1cm);
	  \coordinate [label=below:\textcolor{black} {$5$}] (x3) at (5,-0.1cm);
	  \coordinate [label=below:\textcolor{black} {$6$}] (x3) at (6,-0.1cm);
	  \coordinate [label=below:\textcolor{black} {$7$}] (x3) at (7,-0.1cm);
       \coordinate [label=below:\textcolor{black} {$8$}] (x3) at (8,-0.1cm);
       \coordinate [label=below:\textcolor{black} {$0.9$}] (x3) at (0.1,4.8cm);
       \coordinate [label=below:\textcolor{black} {$0.3$}] (x3) at (0.1,1.8cm);    
       \fill [color=gray] (0.6,0.0) rectangle (1,1.59);
       \fill [color=gray] (1.6,0.0) rectangle (2,1.25);
       \fill [color=gray] (2.6,0.0) rectangle (3,1.25); 
       \fill [color=gray] (3.6,0.0) rectangle (4,3.525);
       \fill [color=gray] (4.6,0.0) rectangle (5,1.82);
       \fill [color=gray] (5.6,0.0) rectangle (6,3.295);
       \fill [color=gray] (6.6,0.0) rectangle (7,1.705);
       \fill [color=gray] (7.6,0.0) rectangle (8,2.045);
       \fill [color=orange] (1,0.0) rectangle (1.4,2.75);
       \fill [color=orange] (2,0.0) rectangle (2.4,0.7);
       \fill [color=orange] (3,0.0) rectangle (3.4,0.7);
       \fill [color=orange] (4,0.0) rectangle (4.4,2.5);
       \fill [color=orange] (5,0.0) rectangle (5.4,1.8);
       \fill [color=orange] (6,0.0) rectangle (6.4,1.15);
       \fill [color=orange] (7,0.0) rectangle (7.4,1.6);
       \fill [color=orange] (8,0.0) rectangle (8.4,1.5);
       \draw[dashed,black] (0.5,1.5)--(9.0,1.5); 
       \draw[dashed,black] (0.5,4.5)--(9.0,4.5); 
       \fill [color=gray]   (7.2,4.0) rectangle (8.2,4.1);
       \fill [color=orange] (7.2,3.6) rectangle (8.2,3.7);
       \coordinate [label=below:\textcolor{black} {$P$}] (x3) at (8.35,4.4cm); 
        \coordinate [label=below:\textcolor{black} {$D$}] (x3) at (8.35,3.9cm);

\end{tikzpicture}
\caption{-Índices de dificultad ($P$) y discriminación ($D$) de cada pregunta.}
\label{Preg}
\end{center}
\end{figure}

El test se puso a punto en dos pasos: primero, confrontando su pertinencia disciplinar con diferentes profesores y segundo, ensayándolo en una muestra de $48$ estudiantes\footnote{ Este test se puede encontrar en el sitio \verb"http://algodefisica.blogspot.com/2012/05/per.html",\\la contraseña de apertura es galileo\verb"$\_$"MUR.}. Se realizó un análisis clásico de test, donde  los indicadores evaluados, de acuerdo con el número de preguntas, fueron el índice de dificultad $(P)$, el índice de discriminación de ítem $(D)$ y el índice global delta de Ferguson ($\delta$) \cite{BEMA}. El índice de dificultad promedio de la prueba fue $0.42$ que está en el intervalo de valides $[0.3,0.9]$; el promedio del índice de discriminación fue $0.34$ que es mayor que el mínimo aceptado $0.3$ y el delta de Ferguson  fue de $0.93$ también mayor que el límite inferior $0.9$.  Los resultados de índice de dificultad y discriminación por ítem se muestran en la gráfica de la Fig.\ref{Preg}. Lo anterior muestra que este test es fiable para usar como indagador en la investigación referida a esta población.

Dado que el test de investigación debía aplicarse $10$ veces consecutivas y existía la posibilidad de que los estudiantes memorizaran las respuestas y compartieran información fuera del aula se construyeron $10$ formas diferentes del test con las preguntas y respuestas distribuidas al azar. Así mismo, para contrarrestar la desmotivación que produce responder la misma prueba de manera rutinaria fue necesario impartir un bono por cada vez que fuese respondido.

Para esta investigación se planteó aplicar  secuencias didácticas de tutoriales, que es una instrucción basada en el aprendizaje activo de la física y desarrollada  desde hace casi tres décadas por Lillian McDermott y sus colaboradores en la Universidad de Washington \cite{LMC1,LMC2} , este tipo de instrucción está fundamentada en las características de los estudiantes que incluyen tanto sus preconceptos como sus modelos mentales. Los tutoriales buscan  desarrollar la comprensión conceptual y el razonamiento físico cualitativo, partiendo de situaciones naturales para el estudiante, vistas desde sus preconceptos, y mediante el conflicto cognitivo materializado a través del  dialogo socrático el profesor  va acercando al estudiante a los  conceptos propios de la física. La implementación de esta instrucción  requiere unos procedimientos concretos y fundamentales: primero aplicar un ejercicio corto en cada clase, que busca por un lado aclararle a los estudiantes los contenidos a estudiar y por otro mostrar a los profesores los problemas de aprendizaje que los  estudiantes tienen en ese instante; segundo que los  estudiantes mediante trabajo colaborativo  y la dirección del profesor realicen las actividades propuestas en el tutorial; tercero que los estudiantes realicen los ejercicios extra clase,  que estos sean retroalimentados y estén en el contexto del tutorial \cite{Julio}.
\begin{figure*}[!ht]
\begin{center}
\includegraphics[width=14.0cm,height=8.5cm,angle=0]{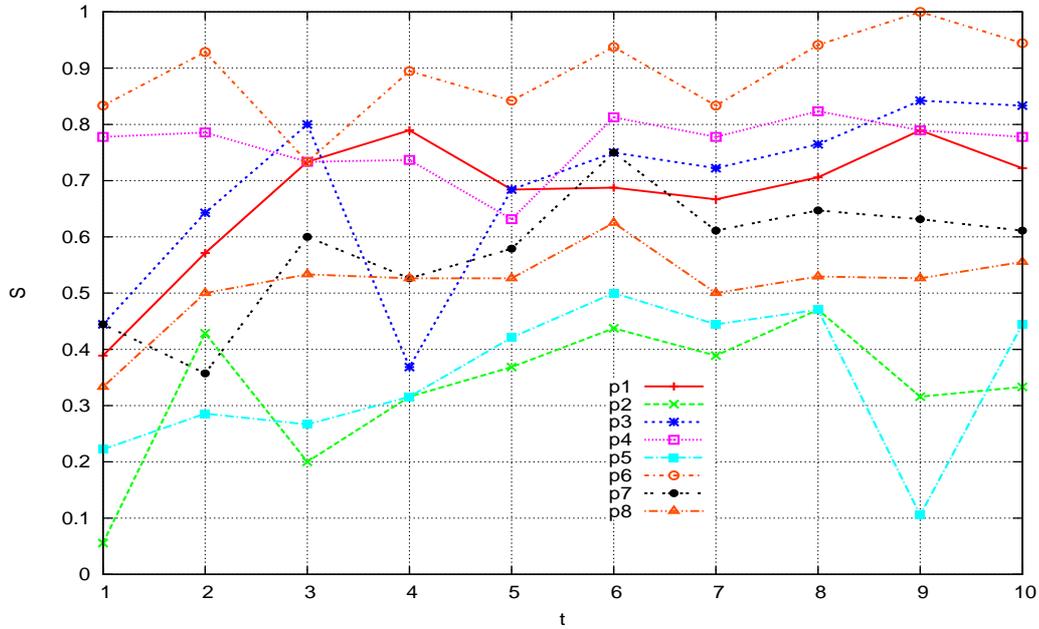}
\caption{-Desempeño $S$ vs tiempo $t$}
\label{CURAP}
\end{center}
\end{figure*}

Para materializar esta  metodología en la investigación se organizó en cada sección de clase grupos de trabajo de cuatro estudiantes cada uno con su propio tutorial donde se discutía las situaciones planteadas en torno a la interpretación gráfica del movimiento rectilíneo y con sus casos particulares uniforme y uniforme acelerado,  en el transcurso de la clase el profesor se paseaba por cada grupo buscando un diálogo socrático entre los estudiantes del grupo y efectuando  algunas explicaciones magistrales, en ciertas  secciones hacia el final de la clase el profesor realizaba un análisis general que terminaba con la solución de los problemas discutidos, finalmente  se dejaba como trabajo extra clase las actividades planteadas en el tutorial y se asignaba una bonificación mediante el sellado de los apuntes. También, se realizaron varias actividades de laboratorio entre las que se destacaron el estudio experimental de las gráficas de posición vs tiempo de cada estudiante en una carrera de $10m$, esta práctica tomo como  material  un registrador simple de tiempo y una cinta de papel, la práctica  permitió  elaborar un poster por estudiante que produjo  una discusión general de los conceptos gráficos del movimiento rectilíneo.  

Sin embargo, se observó que la discusión de los contenidos físicos se  hizo  difícil de desarrollar en el aula debido a la falta de concentración por parte de los  estudiantes, pues por lo general se mantenía  la discusión y confrontación de ideas en un grupo particular mientras el profesor estaba en interacción con tal grupo de lo contrario se limitaban  a completar las actividades del tutorial sin mucho análisis. Así mismo, las actividades extra clase eran realizadas a medias, completadas en último momento, copiadas entre estudiantes  y de libros sin mayor análisis. En general se observó en los estudiantes, poco interés, inmediatez y rápido olvido,  de manera que el compromiso intelectual cargado de crítica, creatividad y autonomía no se logró del todo ni dentro ni fuera del aula. 

Durante el transcurso de la investigación la intencionalidad de respuesta correcta no se mantuvo fija, pues la aplicación estuvo inmersa dentro de las evaluaciones tradicionales y en particular en la aplicación $9$ se realizó el examen parcial. De igual modo, el número de estudiantes no se mantuvo constante debido a la inasistencia a clase y el incorrecto diligenciamiento del test, pues no se escribió el código de la prueba asignada y dado que había diez temas distintos no había manera de identificar con la hoja de respuesta que tema le correspondía.  Así que la asistencia efectiva es la mostrada en la tabla \ref{tab2}. No obstante, es de notar que este hecho  no afecta de ninguna manera la aplicación del DEAF ni las curvas de desempeño del curso.

\begin{table}[!htp]
\caption{Asitencia}\label{tab2} 
\vspace{2mm}
\centering
\begin{tabular}{|c|c||c|c|}
  \hline
$t$ & $N_e$ & $t$ &$N_e$ \\ \hline \hline
 1  &  18   &  6   & 16            \\ 
 2  &  14   &  7   & 18          \\ 
 3  &  15   &  8   & 17        \\ 
 4  &  19   &  9   & 19      \\ 
 5  &  19   &  10  & 19      \\  \hline  
\end{tabular}
\end{table}

\begin{figure}[ht]
\begin{center}
\begin{tikzpicture}[scale=1.6,decoration=Koch snowflake,x=1cm,y=1cm]
     \draw[dashed,black] (1,1)--(1,3);
     \draw[dashed,black] (1,3)--(5,5);
     \draw[dashed,black] (5,5)--(2.5,0);
     \draw[dashed,black] (2.5,0)--(1,1);
	\draw[dashed,black] (1,1)--(5,5);
     \draw[-latex,rounded corners,red,line width=0.7pt]
          (1.94,0.90)--(2.86,1.35)--(3.67,2.42)--(3.95,2.89)--(3.42,2.38)--(3.44,2.06)
           --(3.33,1.90)--(3.53,2.21)--(3.95,2.89)--(3.61,2.43);
      \draw[red] (1.94,0.90) circle (0.25ex);
      \draw[red] (2.86,1.35) circle (0.25ex);
      \draw[red] (3.67,2.42) circle (0.25ex);
      \draw[red] (3.95,2.89) circle (0.25ex);
      \draw[red] (3.42,2.38) circle (0.25ex);
      \draw[red] (3.44,2.06) circle (0.25ex);
      \draw[red] (3.33,1.90) circle (0.25ex);
      \draw[red] (3.53,2.21) circle (0.25ex);
      \draw[red] (3.95,2.89) circle (0.25ex);
      \draw[red] (3.61,2.43) circle (0.25ex);
      \coordinate [label=below:\textcolor{black} {$p_1$}]   (x3) at (2.86,1.35cm);
     \draw[-latex,rounded corners,green,line width=0.5pt]
         (0.28,1.16)--(2.14,0.83)--(1.00,0.37)--(1.58,0.57)--(1.84,0.70)--(2.19,1.12)--
         (1.94,0.93)--(2.35,1.04)--(1.58,0.91)--(1.67,1.01);
                      
	\draw[green](0.28,1.16) circle (0.25ex); 
	\draw[green](2.14,0.83) circle (0.25ex); 
	\draw[green](1.00,0.37) circle (0.25ex); 
	\draw[green](1.58,0.57) circle (0.25ex); 
	\draw[green](1.84,0.70) circle (0.25ex); 
	\draw[green](2.19,1.12) circle (0.25ex); 
	\draw[green](1.94,0.93) circle (0.25ex); 
	\draw[green](2.35,1.04) circle (0.25ex); 
	\draw[green](1.58,0.91) circle (0.25ex); 
	\draw[green](1.67,1.01) circle (0.25ex); 
     \coordinate [label=below:\textcolor{black} {$p_2$}]   (x3) at (0.28,1.16cm);
      
 \draw[-latex,rounded corners,blue,line width=0.5pt]     
      (4.17,3.21)--(4.64,4.17)--(3.67,2.47)--(4.47,3.83)--(4.21,3.29)--(4.69,4.30)--(4.17,3.25)--
      (4.71,4.33)--(5.00,5.00)--(4.72,4.37);
      
	\draw[blue](4.17,3.21)circle (0.25ex); 
	\draw[blue](4.64,4.17)circle (0.25ex); 
	\draw[blue](3.67,2.47)circle (0.25ex); 
	\draw[blue](4.47,3.83)circle (0.25ex); 
	\draw[blue](4.21,3.29)circle (0.25ex); 
	\draw[blue](4.69,4.30)circle (0.25ex); 
	\draw[blue](4.17,3.25)circle (0.25ex); 
	\draw[blue](4.71,4.33)circle (0.25ex); 
	\draw[blue](5.00,5.00)circle (0.25ex); 
	\draw[blue](4.72,4.37)circle (0.25ex); 
     \coordinate [label=below:\textcolor{black} {$p_6$}]   (x3) at (5.2,5.0cm);            
\end{tikzpicture}
\caption{-Diagrama de evolución de las preguntas $1,2$ y $3$}
\label{Trayec2}
\end{center}
\end{figure}
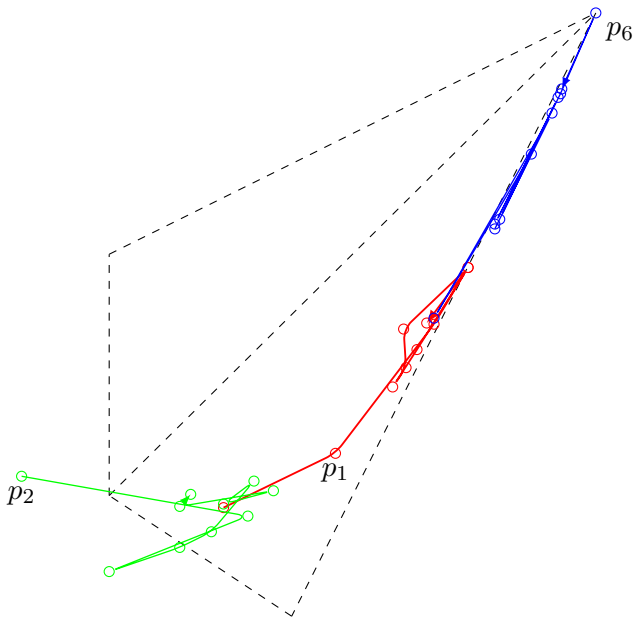

En la Fig.\ref{CURAP} se muestran las curvas típicas de desempeño, cada curva corresponde al desempeño de la población  durante las $10$ aplicaciones consecutivas. Se observa que en general las curvas son muy planas, lo que significa que no hubo mayor cambio en el rendimiento, es decir el desempeño de los estudiantes sólo  mejora un poco pero después decaen y en general oscilan de clase en clase. En general estas curvas de rendimiento resultan similares a las reportadas por \cite{Quepasa}. De la gráfica en la Fig.\ref{CURAP} se infiere la representatividad de las preguntas $1$, $2$ y $6$ las cuales se examinan en cuanto a su evolución mediante el DEAF, ver Fig. \ref{Trayec2}.

La pregunta $(1)$ muestra una evolución pequeña con baja homogeneidad y puntaje medianamente  alto que alcanza el máximo en el penúltimo intento y luego retrocede a un menor puntaje y baja homogeneidad. Dado que esta pregunta busca conocer si el estudiante interpreta la pendiente como la velocidad en una gráfica de posición vs tiempo se puede afirmar que la población en estudio sólo alcanzó medianamente este concepto y que además es oscilante sin mucho parecido entre sus ideas alternativas.  

La pregunta $(2)$ es un ejemplo típico de esos conceptos que no logran evolucionar sino que se quedan oscilando entre lo medianamente claro un día y al siguiente de nuevo confuso. Esta pregunta indaga por la interpretación gráfica de la velocidad vs tiempo desde una perspectiva narrativa, esto indica que las  acciones tendientes a buscar que el estudiante pinte una gráfica a partir de un enunciado literal han fallado con esta metodología. 

En la pregunta $(3)$ se indaga por el desplazamiento visto desde la gráfica de velocidad vs tiempo. Nótese que hay plena evolución hasta alcanzar el máximo, pero no se mantiene sino que retorna a puntos anteriores. En verdad no hay un aprendizaje definitivo del concepto ya que además la evolución se produce algo retirada de la semirrecta central y esto indica que se acercan escogiendo respuestas alternas al azar y no por un descarte que el conocimiento físico de este concepto  debe permitir.      

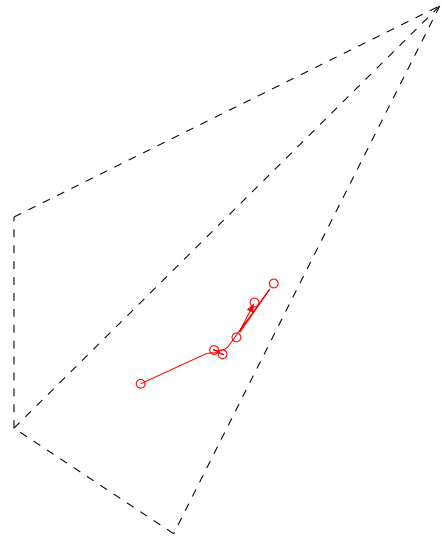
\begin{figure}[ht]
\begin{center}
\begin{tikzpicture}[scale=1.4,decoration=Koch snowflake,x=1cm,y=1cm]
     \draw[dashed,black] (1,1)--(1,3);
     \draw[dashed,black] (1,3)--(5,5);
     \draw[dashed,black] (5,5)--(2.5,0);
     \draw[dashed,black] (2.5,0)--(1,1);
	\draw[dashed,black] (1,1)--(5,5);
     \draw[-latex,rounded corners,red,line width=0.3pt]
        (2.19,1.42)--(2.88,1.74)--(2.96,1.70)--(3.44,2.37)--(3.09,1.86)--(3.26,2.19);
 	
 	\draw[red](2.19,1.42)circle (0.25ex);
 	\draw[red](2.88,1.74)circle (0.25ex);
 	\draw[red](2.96,1.70)circle (0.25ex);
 	\draw[red](3.44,2.37)circle (0.25ex);
 	\draw[red](3.09,1.86)circle (0.25ex);
 	\draw[red](3.26,2.19)circle (0.25ex);

\end{tikzpicture}
\caption{-DEAF aplicado sobre la prueba general}
\label{Trayec3}
\end{center}
\end{figure}

Al aplicar el DEAF a la prueba completa se buscó indagar por el pensamiento físico de los estudiantes en cuanto a la interpretación gráfica del MUR se refiere. Este resultado se muestra en la Fig.\ref{Trayec3}. Nótese que en general se encuentra una leve evolución con retrocesos y oscilaciones y lejos de la semi-recta central, el punto de máxima evolución corresponde a  $(Smax,Hmax)\approx (0.75,0.75)$ alcanzado en la $9$ aplicación que coincide con la fecha de examen, esto significa que la metodología no fue lo suficientemente idónea para  mantener un crecimiento continuo y efectivo del aprendizaje de este tema particular de la física introductoria.  

\section{Conclusiones}

Se mostró el diagrama de evolución de aprendizaje de la física, DEAF, como una técnica que permite estudiar la evolución del aprendizaje de conceptos concretos de física elemental durante algún periodo tiempo. Se explico las ventajas que esta técnica tiene, pues: cuantifica la evolución ya sea por conceptos particulares o de manera global permitiendo intervenir en tiempo de ejecución para cambiar aspectos metodológicos y mejorar la instrucción;  permite observar si la conceptualización correcta de los estudiantes  se mantiene en el tiempo o no; permite saber si los estudiantes responde al azar o con algún conocimiento disciplinar, da elementos para juzgar la efectividad de instrucción y muestra una manera de conocer la homogeneidad disciplinar de los estudiantes.   Adicionalmente, esta técnica plantea algunos retos a la comunidad PER ya que crea la necesidad  de diseñar metodologías de aplicación de test indistinguibles desde el punto de vista físico pero completamente diferentes a la percepción de los estudiantes, para ser aplicados  múltiples veces sin causar cansancio ni generar recuerdos y así disponer de instrumentos que fortalezcan el estudio de la evolución del aprendizaje  de la física. 

El estudio realizado con DEAF aplicado en estudiantes de ingeniería de la Universidad Central arrojó  resultados que no muestran la metodología de tutoriales  como la más efectiva para este grupo. Esto quizá se deba a la poca generación de interés de los estudiantes por este tipo de temas; así como a la imposibilidad de desarrollar únicamente el tema de la interpretación gráfica del movimiento unidimensional durante todo el semestre; también pudo afectar la falta de un acompañamiento orientado al tema especifico en las actividades extra clase mediante monitores o profesores conocedores de la metodología y sobre todo,  la poca efectividad lograda se debe a que no se han pensado ni desarrollado estrategias que contrarresten la tendencia cultural de la inmediatez y el análisis superfluo. Es decir, el trabajo logró falsear la hipótesis planteada.      

Dos perspectivas de investigación futura proyecta esta investigación: la primera consiste en aplicar DEAF a la metodología de intrusión entre pares y observar la relación  la evolución de conceptos concretos durante una clase, segundo observar la evolución de conceptos simples durante todo un programa a lo largo de $5$ años y obtener conclusiones sobre las metodologías a partir del DEAF.    

\section*{Agradecimientos}

Los autores agradecen a los estudiantes de física I que participaron en la investigación durante los periodos académicos $2011-2$ y $2012-1$; a los  profesores que revisaron el test y prestaron sus cursos para ponerlo a prueba;  al Departamento de Ciencias Naturales de la Universidad  Central por el apoyo y el tiempo asignado a la investigación y al CICATA del IPN de México por su continua colaboración. 

\renewcommand{\refname}{Referencias}

\end{document}